\documentclass[]{spie}  

\usepackage{amsmath,amsfonts,amssymb}
\usepackage{graphicx}
\usepackage[colorlinks=true, allcolors=blue]{hyperref}

\title{Walking with SOXS towards the transient sky}

\author[a]{P.~Schipani}
\author[b]{S.~Campana}
\author[c]{R.~Claudi}
\author[b]{M.~Aliverti}
\author[a]{A.~Baruffolo}
\author[d]{S.~Ben-Ami}
\author[a]{G.~Capasso}
\author[a]{M.~Colapietro}
\author[e]{R.~Cosentino}
\author[f]{F.~D'Alessio}
\author[b]{P.~D'Avanzo}
\author[b]{M.~Genoni}
\author[d]{O.~Hershko}
\author[g,h]{H.~Kuncarayakti}
\author[b]{M.~Landoni}
\author[k]{M.~Munari}
\author[m,n]{G.~Pignata}
\author[c]{K.~Radhakrishnan}
\author[c]{D.~Ricci}
\author[o]{A.~Rubin}
\author[p]{S.~Scuderi}
\author[f]{F.~Vitali}
\author[q]{D.~Young}
\author[o]{M.~Accardo}
\author[r]{J.~Achrén}
\author[s]{J.~A.~Araiza-Duran}
\author[t]{I.~Arcavi}
\author[b]{L.~Asquini}
\author[c,u]{F.~Battaini}
\author[d]{A.~Bichkovsky}
\author[d]{A.~Brucalassi}
\author[d]{R.~Bruch}
\author[c]{L.~Cabona}
\author[c]{E.~Cappellaro}
\author[a]{M.~Della~Valle}
\author[c]{S.~Di~Filippo}
\author[k]{R.~Di~Benedetto}
\author[a]{S.~D'Orsi}
\author[d]{A.~Gal-Yam}
\author[e]{M.~Hernandez}
\author[o]{D.~Ives}
\author[o]{H.-U.~K\"aufl}
\author[h,g]{J.~Kotilainen}
\author[v]{G.~Li~Causi}
\author[c]{L.~Lessio}
\author[a]{L.~Marty}
\author[g]{S.~Mattila}
\author[o]{L.~Mehrgan}
\author[r]{L.~Pasquini}
\author[w]{E.~Pompei}
\author[d]{M.~Rappaport}
\author[b]{M.~Riva}
\author[c]{B.~Salasnich}
\author[a]{S.~Savarese}
\author[w]{I.~Saviane}
\author[o]{M.~Sch\"oller}
\author[w]{A.~Silber}
\author[x]{S.~Smartt}
\author[a]{R.~Zanmar~Sanchez}
\author[y]{M.~Stritzinger}
\author[z]{A.~Sulich}
\author[e]{H.~Ventura}

\affil[a]{INAF - Osservatorio Astronomico di Capodimonte, Sal. Moiariello 16, I-80131, Naples, Italy }
\affil[b]{INAF - Osservatorio Astronomico di Brera, Via Bianchi 46, I-23807, Merate, Italy }
\affil[c]{INAF - Osservatorio Astronomico di Padova, Vicolo dell’Osservatorio 5, I-35122, Padua, Italy }
\affil[d]{Weizmann Institute of Science, Herzl St 234, Rehovot, 7610001, Israel }
\affil[e]{FGG-INAF, TNG, Rambla J.A. Fernández Pérez 7, E-38712 Breña Baja (TF), Spain }
\affil[f]{INAF - Osservatorio Astronomico di Roma, Via Frascati 33, I-00078 M. Porzio Catone, Italy }
\affil[g]{Tuorla Observatory, Dept. of Physics and Astronomy, University of Turku,  FI-20014, Finland }
\affil[h]{Finnish Centre for Astronomy with ESO (FINCA), FI-20014 University of Turku, Finland }
\affil[k]{INAF - Osservatorio Astrofisico di Catania, Via S. Sofia 78, I-95123 Catania, Italy }
\affil[m]{Instituto de Alta Investigación, Universidad de Tarapacá, Casilla 7D, Arica, Chile }
\affil[n]{MAS, Nuncio Monseñor Sotero Sanz 100, Providencia, Santiago, Chile }
\affil[o]{ESO, Karl Schwarzschild Strasse 2, D-85748, Garching bei München, Germany }
\affil[p]{INAF - Istituto di Astrofisica Spaziale e Fisica Cosmica, Via Corti 12, I-20133, Milano, Italy}
\affil[q]{Astrophysics Research Centre, Queen's University Belfast, Belfast, BT7 1NN, UK }
\affil[r]{Incident Angle Oy, Capsiankatu 4 A 29, FI-20320 Turku, Finland }
\affil[s]{INAF - Osservatorio Astrofisico di Arcetri, Largo E. Fermi 5, I-50125, Firenze, Italy}
\affil[t]{Tel Aviv University, Department of Astrophysics, 69978 Tel Aviv, Israel }
\affil[u]{Dipartimento di Fisica e Astronomia ``G. Galilei'', Universit\`a di Padova, Italy }
\affil[v]{INAF - Istituto di Astrofisica e Planetologia Spaziali, I-00133 Rome, Italy}
\affil[w]{ESO, Alonso de Cordova 3107, Vitacura, Casilla 19001, Santiago de Chile 19, Chile}
\affil[x]{University of Oxford, Keble Road, Oxford OX1 3RH, UK}
\affil[y]{Aarhus University, Ny Munkegade 120, D-8000 Aarhus, Denmark}
\affil[z]{INAF - Osservatorio Astronomico di Trieste, Via G.B. Tiepolo 11, I-34143 Trieste, Italy }

\authorinfo{Send correspondence to: pietro.schipani@inaf.it}

\begin{document} 
\maketitle

\begin{abstract}
SOXS (Son Of X-Shooter) is the new ESO instrument that is going to be installed on the 3.58-m New Technology Telescope at the La Silla Observatory. 
SOXS is a single object spectrograph offering a wide simultaneous spectral coverage from U- to H-band. Although such an instrument may have 
potentially a large variety of applications, the consortium designed it with a clear science case: it is going to provide the spectroscopic counterparts to the 
ongoing and upcoming imaging surveys, becoming one of the main  follow-up instruments in the Southern hemisphere for the classification and 
characterization of transients. 
The NTT+SOXS system is specialized to observe all transients and variable sources discovered by imaging surveys with a  flexible schedule 
maintained by the consortium, based on a remote scheduler which will interface with the observatory software infrastructure. SOXS is realized timely 
to be highly synergic with transients discovery machines like the Vera C. Rubin Observatory. The instrument has been integrated and tested in Italy,
collecting and assembling subsystems coming from all partners spread over six countries in three continents. The first preparatory activities in Chile 
have been completed at the telescope. This article gives an updated status of the project before the shipping of the instrument to Chile.
\end{abstract}


\begin{figure} [ht]
\begin{center}
\begin{tabular}{c} 
\includegraphics[height=11.9cm]{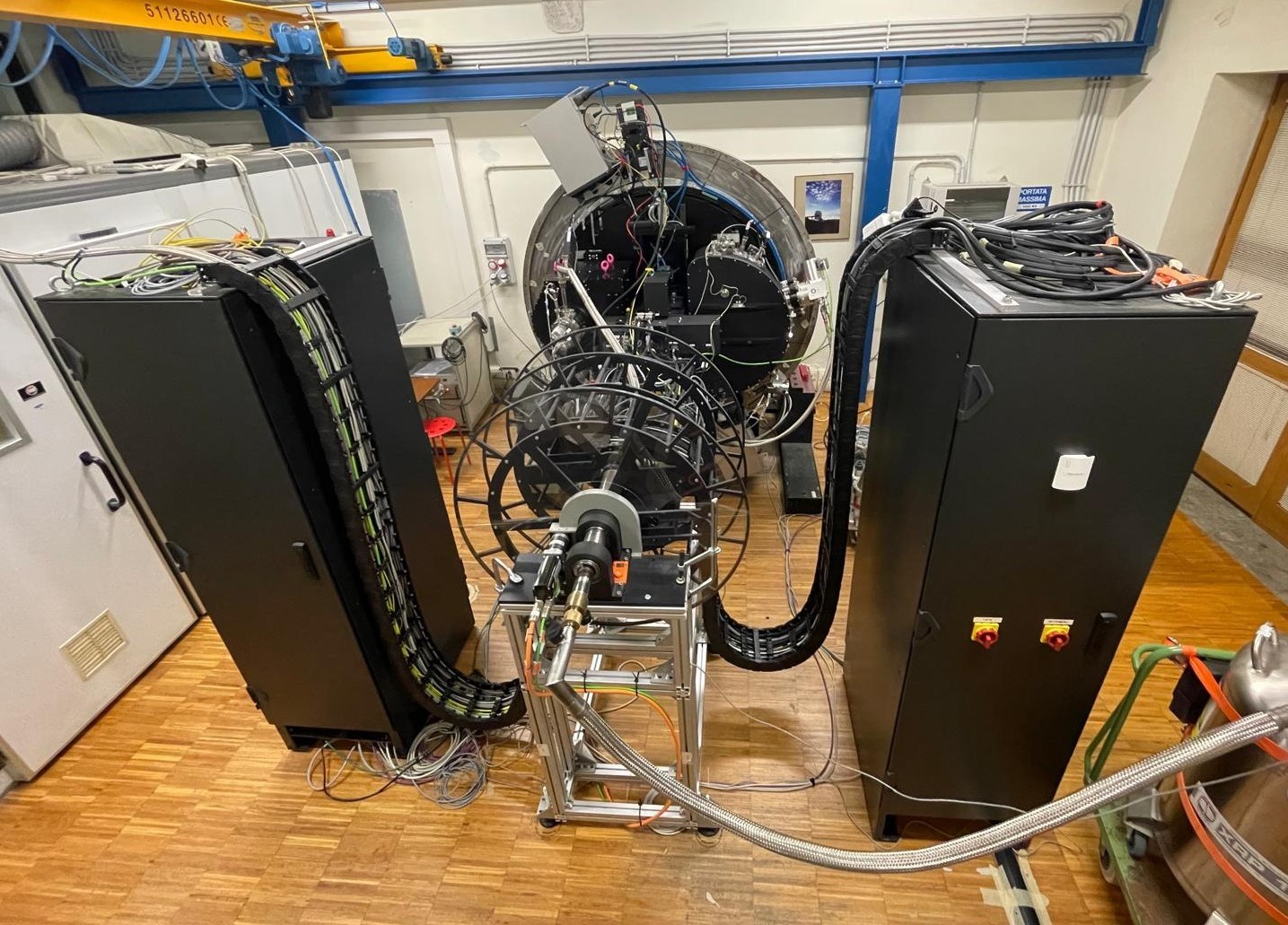}
\end{tabular}
\end{center}
\caption[SOXS] 
{ \label{fig:SOXS} 
The SOXS instrument fully assembled in the integration hall of the Padova Astronomical Observatory. It is mounted on the NTT 
simulator flange, providing the same interfaces of the real telescope at the Nasmyth focus. The instrument can be rotated and tested in
all gravity conditions. The co-rotator structure is visible in the center with  
two lateral cable-wraps which safely support the cables during the rotation. The two black cabinets on the left and right host most of the 
control electronics. The only missing part in this picture is the Nasmyth platform, which has already been shipped and installed in Chile.
 }
\end{figure}

\begin{figure} [ht]
\begin{center}
\begin{tabular}{c} 
\includegraphics[height=11cm]{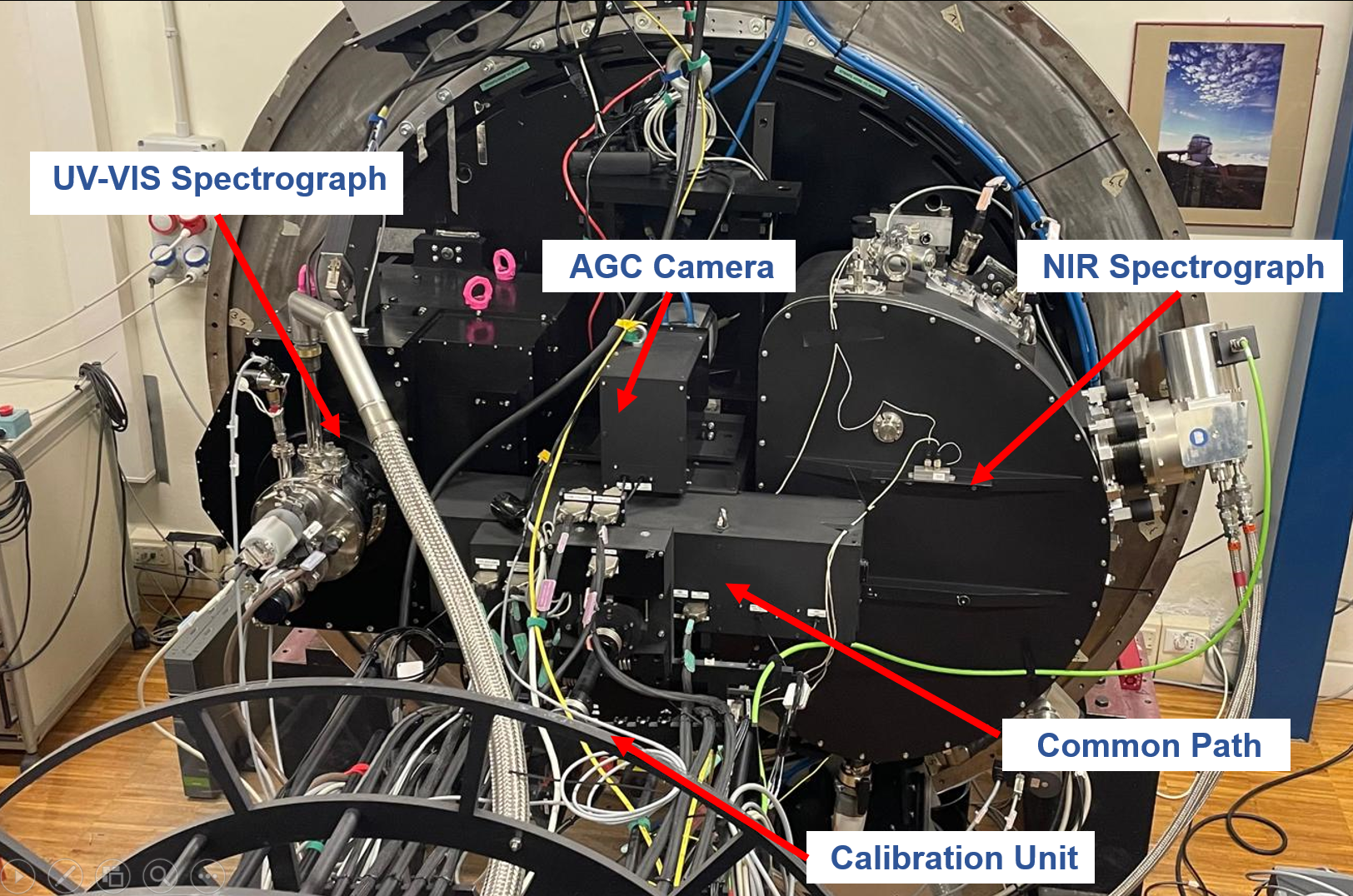}
\end{tabular}
\end{center}
\caption[example1] 
{ \label{fig:Subsys} 
The five main subsystems of SOXS. The Common Path, the heart of the system, is surrounded by the other four: the two spectrographs working
simultaneously on the left (UV-VIS) and right (NIR), the Acquisition and Guiding Camera on the top, the Calibration Unit on the bottom.}
\end{figure}

\begin{figure} [ht]
\begin{center}
\begin{tabular}{c} 
\includegraphics[height=11.9cm]{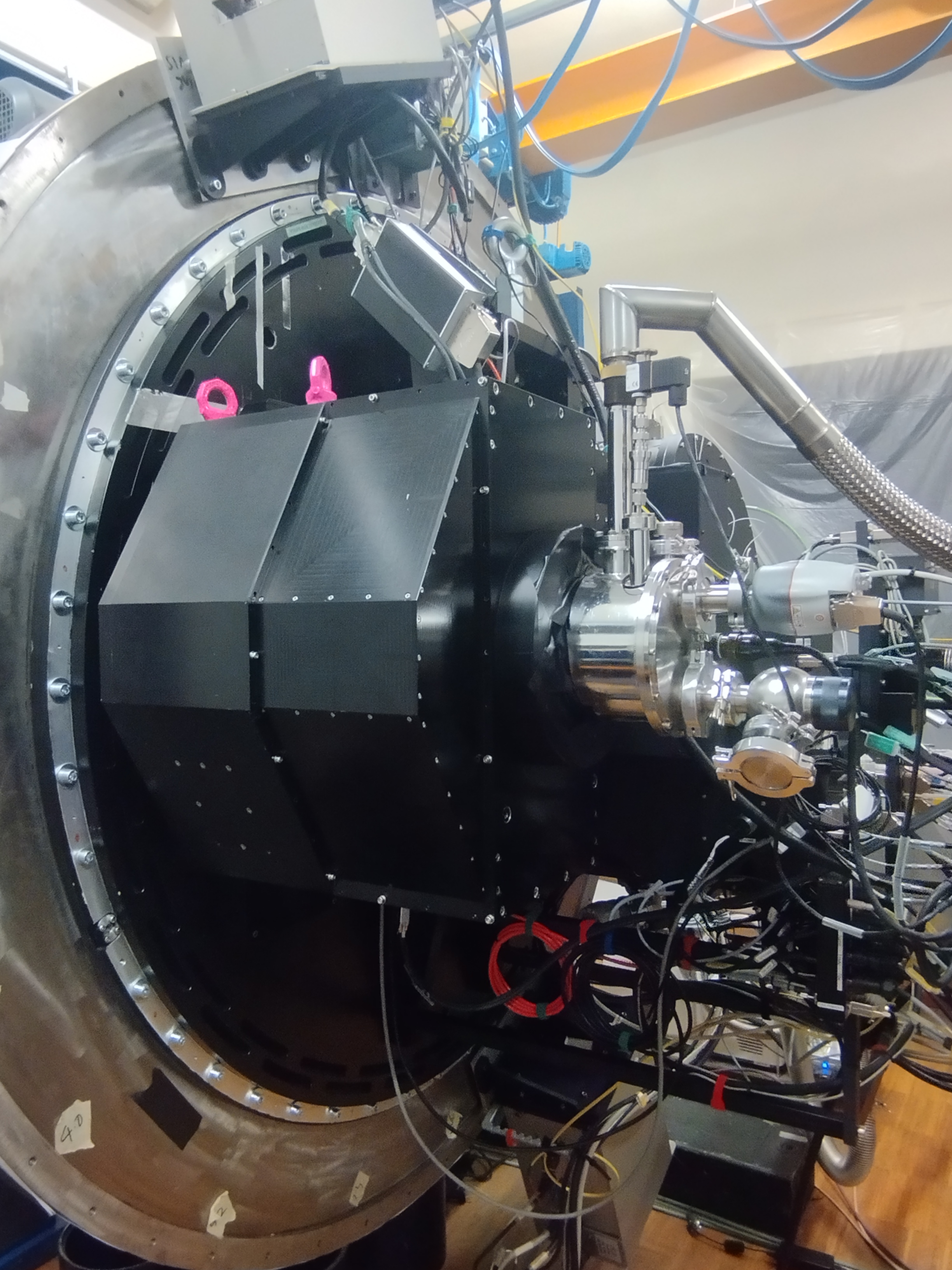}
\end{tabular}
\end{center}
\caption[example1] 
{ \label{fig:left} 
The UV-VIS spectrograph of SOXS. The detector is cooled through a Continuous Flow Cryostat with liquid nitrogen.
 }
\end{figure}

\begin{figure} [ht]
\begin{center}
\begin{tabular}{c} 
\includegraphics[height=5.75cm]{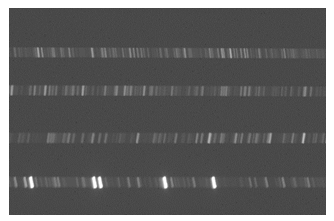}
\includegraphics[height=5.85cm]{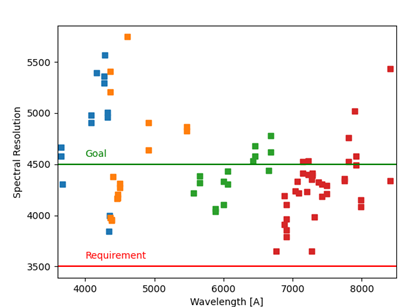}
\end{tabular}
\end{center}
\caption[example1] 
{ \label{fig:UVVISRes} 
Left: UVVIS four traces when fed through the CP using SoXS CBX arclamps. Right: Measured spectral resolution, shown average resolution 
of R$\sim$4500, meeting all system requirements, and achieving goals in most cased – see text for further details.}
\end{figure}

\begin{figure} [ht]
\begin{center}
\begin{tabular}{c} 
\includegraphics[height=11.9cm]{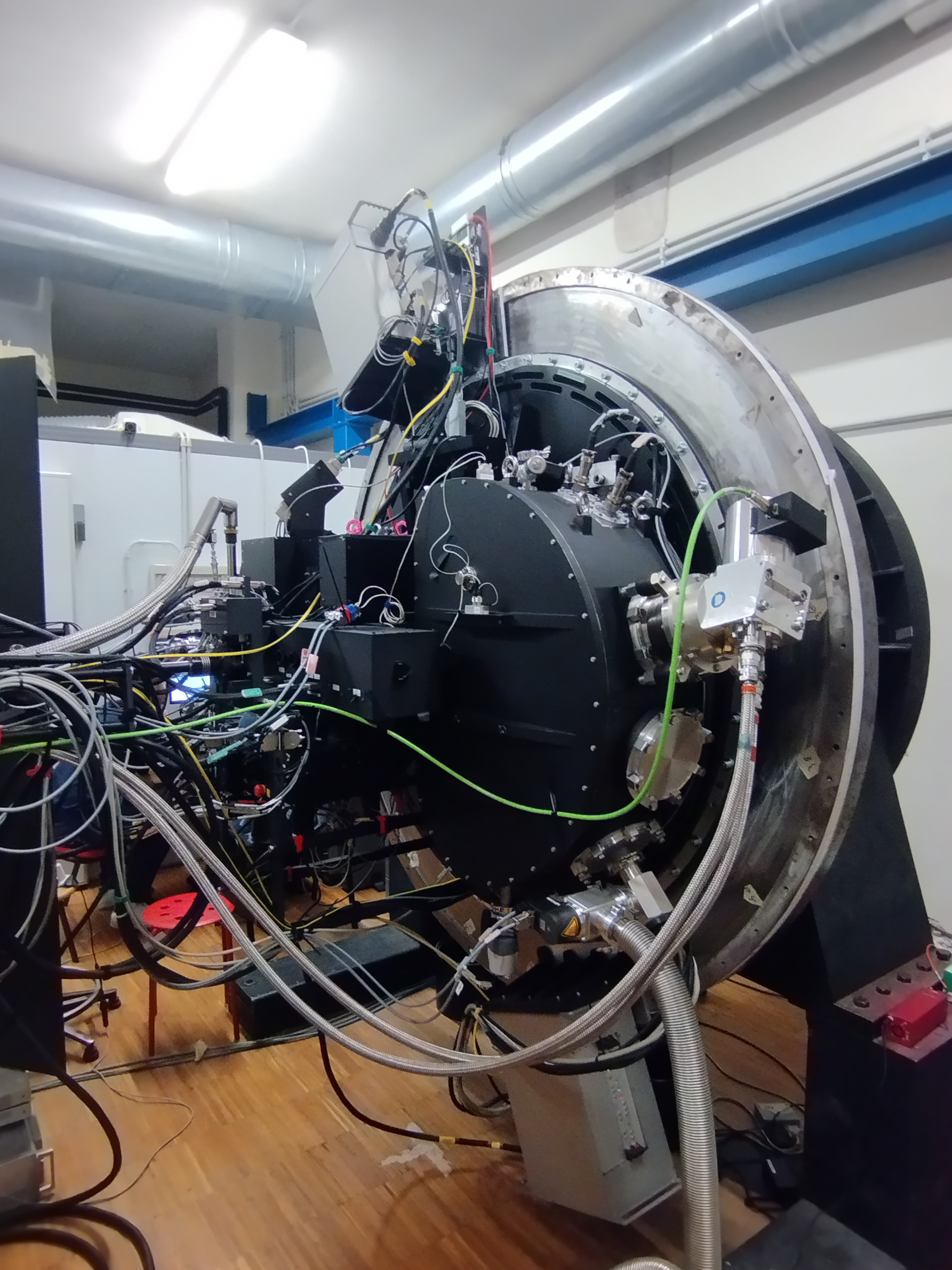}
\end{tabular}
\end{center}
\caption[example1] 
{ \label{fig:right} 
The NIR spectrograph of SOXS. The cryostat is cooled with a CryoCooler.
 }
\end{figure}

\begin{figure} [ht]
\begin{center}
\begin{tabular}{c} 
\includegraphics[height=10cm]{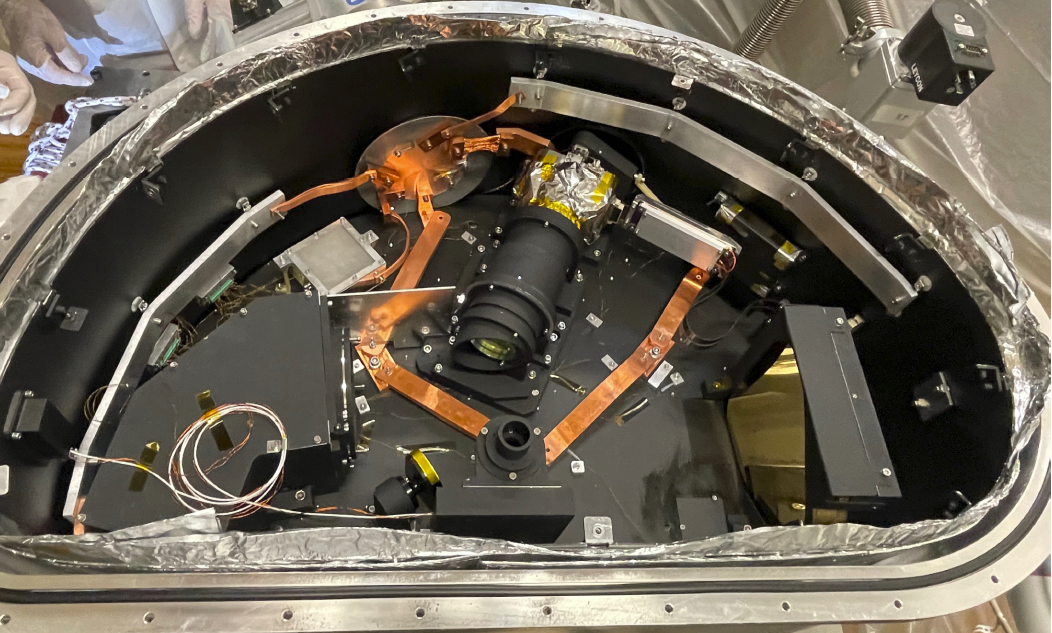}
\end{tabular}
\end{center}
\caption[example1] 
{ \label{fig:NIR_vessel} 
Inner view of the NIR spectrograph. 
 }
\end{figure}

\begin{figure} [ht]
\begin{center}
\begin{tabular}{c} 
\includegraphics[height=6.7cm]{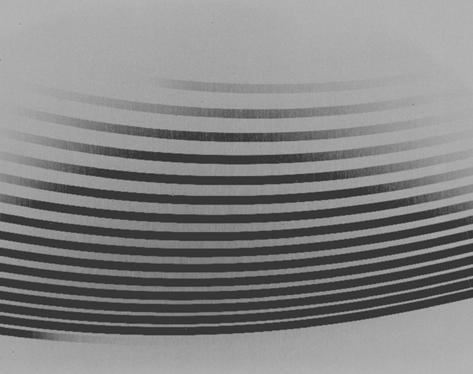}
\includegraphics[height=6.7cm]{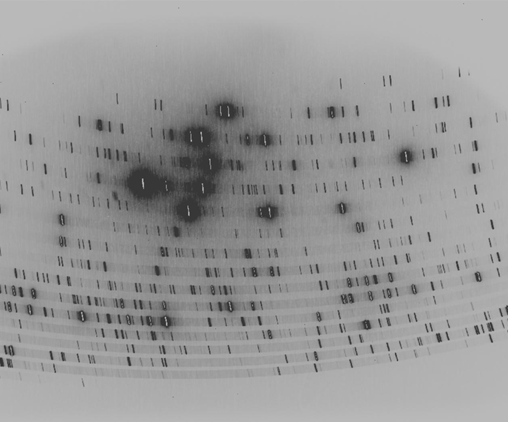}
\end{tabular}
\end{center}
\caption[example1] 
{ \label{fig:NIRSpectra} 
Echelle cross-dispersed spectral format of the NIR spectrograph. Left: using continuum (QTH) lamp. Right: using arc lamps with 0.5 arcsec slit.}
\end{figure}

\begin{figure} [ht]
\begin{center}
\begin{tabular}{c} 
\includegraphics[height=9cm]{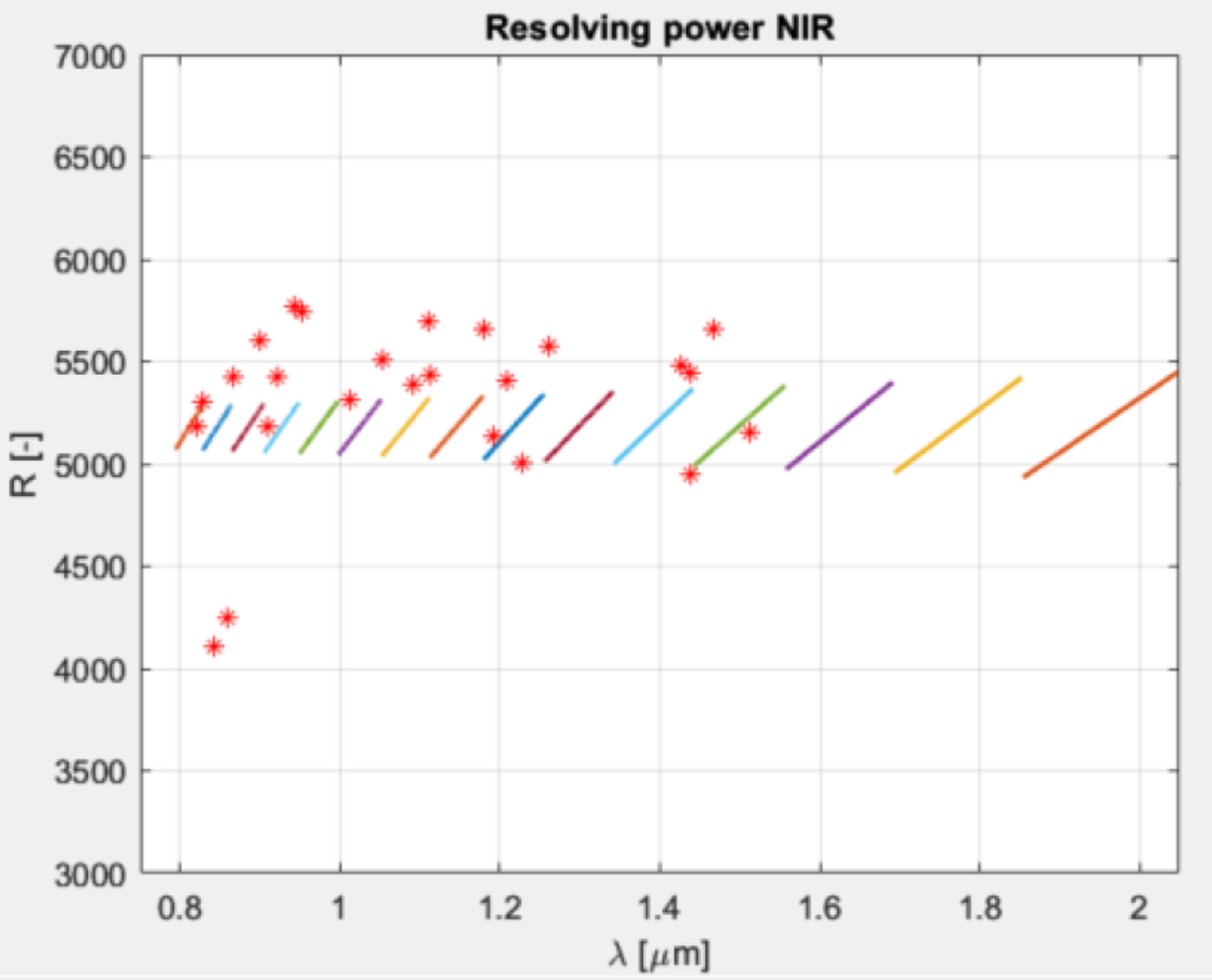}
\end{tabular}
\end{center}
\caption[example1] 
{ \label{fig:NIR_res} 
Measurements of the NIR spectrograph resolution. 
 }
\end{figure}

\begin{figure} [ht]
\begin{center}
\begin{tabular}{c} 
\includegraphics[height=13.5cm]{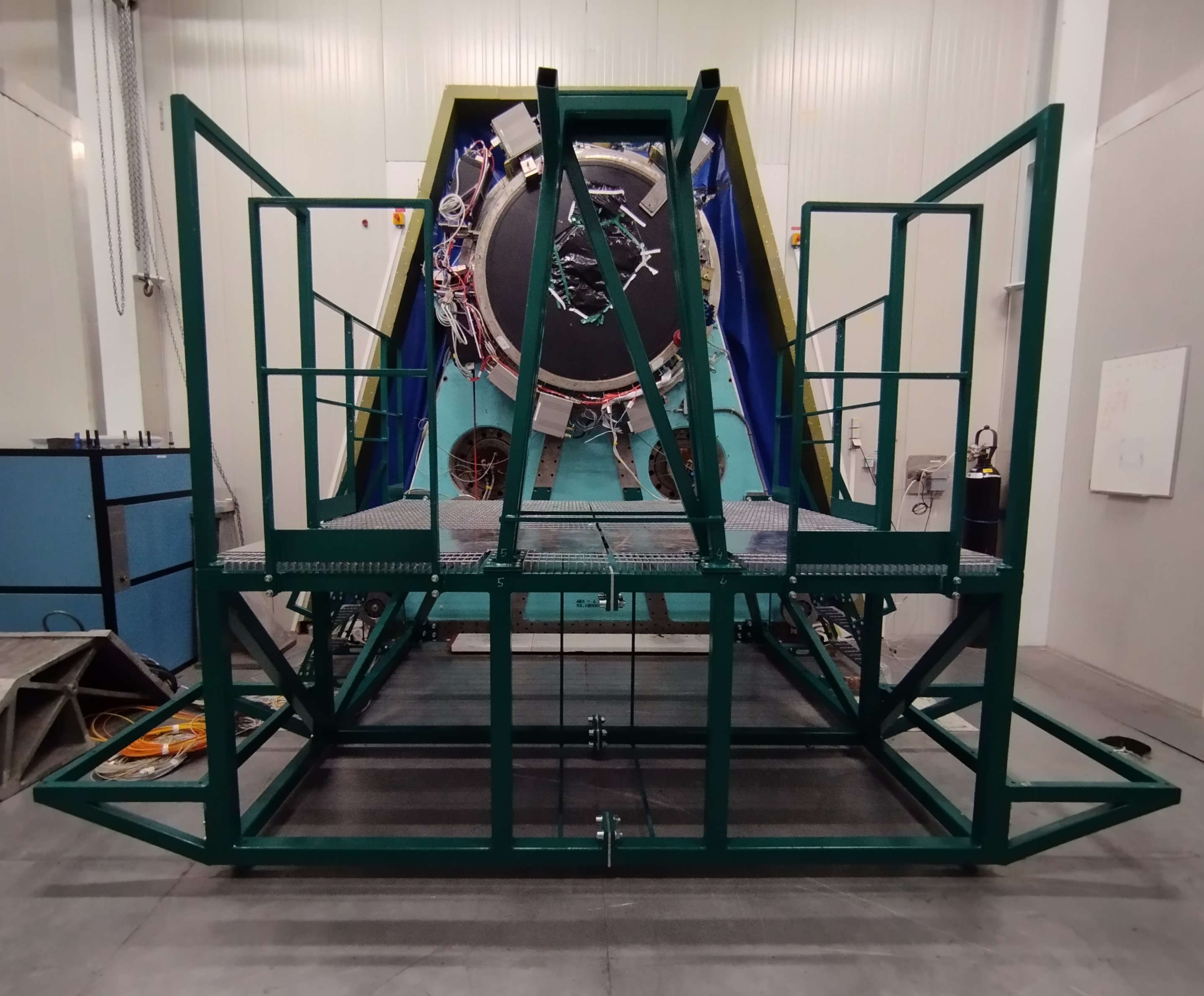}
\end{tabular}
\end{center}
\caption[example1] 
{ \label{fig:platformFront} 
The SOXS platform fully assembled and installed at the NTT in the La Silla Observatory. It will support the co-rotator structure as well as the two
electronic cabinets.
 }
\end{figure}

\section{INTRODUCTION}
\label{sec:intro}  

The SOXS instrument has been introduced in previous contributions, reporting on the status of the project at regular 
intervals\cite{Schipani16,Schipani18,Schipani20,Schipani22}. 
The project originated with an ESO call for new scientific ideas at the NTT telescope. The SOXS consortium participated with a proposal for a new instrument fully dedicated to the spectroscopic follow-up of transient sources, which was selected. The SOXS concept fits with the strategy of dedication of medium class telescopes to specific science cases. In fact, having lot of observing time is a winning factor in many cases, and this is where nowadays the 4-m class telescope can be strategic. The project was officially adopted and kicked-off in 2016, receiving financial support from the partners. Afterwards, it has went through all the regular steps of ESO projects, having preliminar and final design phases and reviews. The procurement and then the assembly and integration phases started later on, and after several years they are now concluded. 
This paper is about the progress of the project towards the first light of the instrument, that will be in 2025.
More details on each subsystem can be found in a set of related articles\cite{Aliverti18,Aliverti20,Aliverti22,Araiza22,Asquini22,Battaini22,Biondi18,Biondi20,Brucalassi18,Brucalassi20,Capasso18,Colapietro20,Cosentino18,Cosentino20,Cosentino22,Genoni20,Genoni22,Kuncarayakti20,Landoni22,Kalyan22,Ricci18,Ricci20,Rubin18,Rubin20,Sanchez18,Sanchez20,Scuderi22,Vitali18,Vitali20,Vitali22,Young22}.

\section{Scientific case}
\label{sec:science} 
There is a wide expectation for the upcoming golden era in the study of astrophysical transients. In fact, many new facilities are coming out
from ground and space at all wavelenghts with unprecedented capabilities, greatly boosting this kind of science. 

A big step forward will come e.g. 
from the imaging all-sky surveys. In that domain the Vera C. Rubin Observatory is going to be a game-changer, as it will discover a large amount of
new sources every day at an unprecedently high rate. But all these sources would need a spectroscopic follow up to understand what they are, and
there are currently no facilities with the capability of providing spectroscopy for the number of new sources that will continuously come out. 
This is where SOXS wants to play a role: being dedicated to the study of transients, it will be available every night for the classification and
characterization of the most interesting targets selected by the consortium. Members of the collaboration will work every day to the selection of 
the most appropriate sources for the next night, with a highly dynamic scheduling imposed by the full dedication to the time-domain 
and multi-messenger astronomy.

The consortium is organized to deal with all kind of astrophysical transients. A non-exhaustive list of the scientific cases is:

\begin{itemize}
    \item Supernovae
    \item Multimessenger event counterparts (Gravitational Waves, neutrino)
    \item Gamma-Ray Bursts
    \item Fast Radio Bursts
    \item Tidal disruption events        
    \item X-ray binaries
    \item Magnetars
    \item AGN, Blazars
    \item Young Stellar Objects, Stellar Variability, Exoplanets
    \item Asteroids and Comets	
\end{itemize}

\section{Instrument Description}
\label{sec:instrument} 
Fig.~\ref{fig:SOXS} shows SOXS fully assembled at the Astronomical Observatory of Padova, where the assembly, integration and test of the system have been performed. It is mounted on the NTT Simulator provided by ESO, which reproduces exactly the real interfaces of the telescope.
The instrument is composed of two spectroscopic channels, working simultaneously in the UV-VIS (350-850 nm) and NIR (800-2000 nm) bands. It is also equipped with an Acquisition and Guiding Camera, that is usable also as an imaging channel for photometry, and a Calibration Unit with lamps for calibration in wavelength and flux. The heart of the system is the Common Path, which receives a F/11 beam from the telescope and transmits a F/6.5 beam to the spectrographs through a dicroich and a set of folding mirrors. The Common Path also hosts most of the mechanisms, e.g. to select the light coming from the calibration lamps, deviate the light to the imaging camera, counter-rotate the atmospheric dispersion corrector prisms, only for visible wavelengths, compensate the flexures with tip-tilt mirrors and so on.  Fig.~\ref{fig:Subsys} highlights the instrument subsystems and their relative positions.

\section{Status}
\label{sec:status} 
This section summarizes the latest results on the three scientific channels, that are the two spectrographs working simulataneously and the imaging camera.

\subsection{UV-VIS Spectrograph}
The UV-VIS spectrograph (Fig.\ref{fig:left}) has gone through extensive assembly, integration and test period during 2023-24. Following assembly of the various optical components,
the science 
detector has been installed and the camera has been aligned to balance image quality and resolution between the four traces. Initial alignment was performed using a 
stand-alone F/6.5 relay with a 50$ \mu$m multimode fiber as a source. Slit alignment was completed using a 3mm liquid fiber and the F/6.5 relay.  
Alignment of the spectrograph to the Common Path is mitigated, as the telescope focus will be adjusted on sky based on the Common Path image plane being coincident 
with the spectrograph object plane.

Using Hg and Kr arclamps, the image quality and spectral resolution has been inspected in a standalone mode, as well as through the 
spectrograph Common Path using the Calibration Unit arc lamps, see Fig.~\ref{fig:UVVISRes} left panel. We derive consistent results between the two modes, with average resolution 
of R=[4830,4560,4480,4350] for 1 arcsec slit width in the four traces of the  spectrograph (u, g ,r ,I respectively), significantly higher 
than the system requirements (R=3500), and either achieving or close to the system goal resolution of R=4500, see Fig.~\ref{fig:UVVISRes} right panel. The 
derived resolution are likely lower limits, as the line width of some of the lines showed below are limiting the spectral resolution, and not 
the system response. Currently, the UV-VIS spectrograph is going through flexure tests, in which we discovered a flexure in the 
spatial direction when completing a rotation on the Nasmyth flange. This residual problem is going to be mitigated by introducing additional supports to the component, which is expected to fix the issue. 

\subsection{NIR Spectrograph}
In the last two years the NIR spectrograph (Fig. \ref{fig:right}) has been fully integrated, aligned and tested. 
All sub-benches were carefully installed and aligned into the optical bench, inside
the D-shape vacuum vessel (Fig.\ref{fig:NIR_vessel}), undergoing mechanical measurements performed with an articulated arm and then optical tests. 
The functional and performance verification tests show that the spectral format, image quality and resolving power, as well as
thermal+dark current noise level are all in agreement with design and compliant with requirements.

Real data in terms of spectral format (Fig.\ref{fig:NIRSpectra} shows the echelle cross-dispersed spectra for continuum and arc lamp sources) and image quality have been compared with simulated frames, generated through the SOXS End-to-End simulator. This approach allowed to confirm the performance were satisfactory.
Flexure and spectral stability tests were executed, coming to the conclusion they are relatively small and repeatable, The  compensation of the flexures is expected to be implemented through a look-up-table mapping the position of the tip-tilt mirror included in the Common Path as a function of the rotator angle.

The Resolving Power estimations were derived from the FWHM values
computed for a set of 25 wavelengths, taking into account their linear dispersion. The outcome of the current analysis is that the resolution for 1 arcsec slit is in agreement with theoretical design values, as shown in Fig.\ref{fig:NIR_res}.

\subsection{AGC Camera}
The Acquisition and Guiding Camera is also considered an additional scientific channel. Thus, its performance have been carefully checked and characterized. The unit acts as a focal reducer, reducing the incoming F/11 beam to a F/3.6. Each pixel of the CCD corresponds to approximately 0.2 arcsec, resulting in a 3.5 arcmin unvignetted field of view. The image quality
specifications is such that over 80$\%$ of the geometrically encircled energy is contained in two pixels in a Field of View with a radius of 3.5 arcmin.

Tests with all filters show the FWHM is in fact contained within a radius of 2 pixels, indicating
satisfactory performance. The results suggest that the system always keeps a high level of precision and focus.
Additionally, the flexure testa revealed that the PSF center of mass shifted by just 1 pixel, on average, during
a full rotation of the flange. This limited displacement proves a good stability and robustness of the system’s alignment under all gravity conditions.

\section{Preliminary activities at the observatory}
\label{sec:preliminary} 
SOXS is designed to replace  the SOFI instrument at one of the Nasmyth foci of the NTT. Thus, in late 2023 ESO decomissioned and removed SOFI, leaving room for the new instrument. The old Nasmyth platform was removed as well, because the SOXS team designed a new lighter platform tailored to the real needs of the new project. This new platform was manufactured in Italy and shipped to Chile in 2023. In November 2023 the SOXS team worked in Chile for the installation of the new platform on the telescope. Fig.~\ref{fig:platformFront} shows the current status in the NTT Nasmyth room, with the new platform waiting for the arrival of the instrument (Fig.~\ref{fig:SOXS}). Of course the visit of the consortium onsite has been extremely useful also to prepare the arrival of the instrument itself, allowing a cross-check of the interfaces at the telescope between the NTT and SOXS.

\section{Conclusions}
\label{sec:concl}
SOXS has been fully assembled in Italy in all its components, merging subsystems provided by different partners of the consortium.
All subsystems have been individually tested at the integration site and, when necessary, improved. The interfaces have been checked and the 
control system has been tuned in parallel.
The preliminary integration in Italy is now completed and the final period of system tuning and tests is ongoing. 
The activities in Chile have been started as well, completing the installation of the Nasmyth platform which will host the instrument in La Silla.
The process of the Preliminary Acceptance in Europe is about to start and the shipment of the instrument to Chile is planned within end of 2024, with 
integration and commissioning in 2025. 

\bibliography{SOXS} 
\bibliographystyle{spiebib} 

\end{document}